\documentclass[prb,twocolumn]{revtex4}
\usepackage{graphicx}
\usepackage{hyperref}

\begin{document}

\title{Valence-bond entanglement and fluctuations in random singlet phases}

\author{Huan Tran\footnote{Current address: Department of Physics, University of Basel, Klingelbergstrasse 82, 4056 Basel, Switzerland} and N. E. Bonesteel}
\affiliation{Department of Physics and National High Magnetic Field Laboratory, Florida State University, Tallahassee, FL 32310, USA}

\date{\today}

\begin{abstract}
The ground state of the uniform antiferromagnetic spin-1/2 Heisenberg chain can be viewed as a strongly fluctuating liquid of valence bonds, while in disordered chains these bonds lock into random singlet states on long length scales. We show that this phenomenon can be studied numerically, even in the case of weak disorder, by calculating the mean value of the number of valence bonds leaving a block of $L$ contiguous spins (the valence-bond entanglement entropy) as well as the fluctuations in this number. These fluctuations show a clear crossover from a small $L$ regime, in which they behave similar to those of the uniform model, to a large $L$ regime in which they saturate in a way consistent with the formation of a random singlet state on long length scales.  A scaling analysis of these fluctuations is used to study the dependence on disorder strength of the length scale characterizing the crossover between these two regimes. Results are obtained for a class of models which include, in addition to the spin-1/2 Heisenberg chain, the uniform and disordered critical 1D transverse-field Ising model and chains of interacting non-Abelian anyons.
\end{abstract}


\maketitle

\section{Introduction}

The set of valence-bond states --- states in which localized spin-1/2 particles are correlated in singlet pairs said to be connected by valence bonds --- provides a useful basis for visualizing singlet ground states of quantum spin systems. For example, the ground state of the uniform one-dimensional nearest-neighbor spin-1/2 antiferromagnetic (AFM) Heisenberg model (the prototypical spin-liquid state\cite{and87}) can be viewed as a strongly fluctuating liquid of valence bonds with a power-law length distribution. This intuitive picture reflects the long-range spin correlations in this state, as well as the existence of gapless excitations created by breaking long bonds.

Valence-bond states also play a key role in describing the physics of random spin-1/2 AFM Heisenberg chains. For these systems, it was shown by Fisher,\cite{fis94} using a real space renormalization group (RSRG) analysis, that on long length scales the ground state is described by a single valence-bond state known as a random singlet state. This single valence-bond state should be viewed as a caricature of the true ground state, which will certainly exhibit bond fluctuations on short length scales. In fact, it is natural to expect that, when measured on these short length scales, a fluctuating random singlet state would be difficult to distinguish from the uniform Heisenberg ground state, particularly in the limit of weak disorder.

In valence-bond Monte Carlo (VBMC) simulations\cite{san05} valence-bond states are used to stochastically sample singlet ground states of quantum spin systems.  One of the appealing features of VBMC is that if one imagines viewing the sampled valence-bond states over many Monte Carlo time steps the resulting ``movie" would correspond closely to the intuitive resonating valence bond picture described above. For random Heisenberg chains (and related models) VBMC should therefore provide a useful method for directly studying the phenomenon of random singlet formation on long length scales, while at the same time capturing the short-range fluctuations which will always be present. 

\begin{figure}[t]
  \begin{center}
    \includegraphics[width=8.25cm]{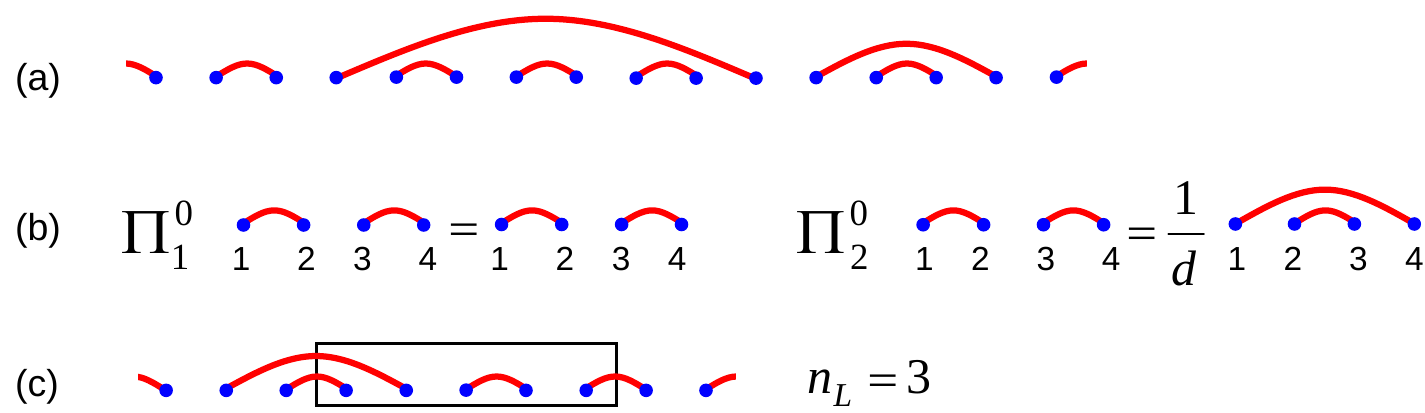}
  \caption{(Color online) (a) A non-crossing valence-bond state.  (b) Action of singlet projection operators $\Pi_i^0$ on a non-crossing valence-bond state.   (c) A non-crossing valence-bond state with a block of $L=5$ sites (region enclosed in the box) for which $n_L$, the number of bonds leaving the block, is 3.} \label{vb}
\end{center}
\end{figure}

With this motivation we have carried out a VBMC study of a class of models which include the uniform and random spin-1/2 AFM Heisenberg chains, as well as models which describe chains of interacting non-Abelian anyons, as special cases. The paper is organized as follows. First, in Sec.~\ref{sec:model}, we define the models and describe their relevant Hilbert spaces. In Sec.~\ref{sec:vbmc}, we give a short review of the VBMC method, and in Sec.~\ref{sec:vbee} present results for the valence-bond entanglement entropy of the uniform and random models.  In Sec.~\ref{sec:vbf} we introduce the valence-bond fluctuations --- a measure of how strongly the valence bonds are fluctuating on a given length scale --- and show that this quantity can be used to provide a clear signature of random singlet state formation. Results of a scaling analysis of the valence-bond fluctuations are then presented in Sec.~\ref{sec:clength} and the paper ends with some conclusions in Sec.~\ref{sec:conclusion}.

\section{Hilbert Space and Model Hamiltonians}\label{sec:model}

To define the class of model Hamiltonians studied here, we first specify the relevant Hilbert spaces on which they act. It is well known that the set of {\it non-crossing} valence-bond states (see Fig.~\ref{vb}(a)) forms a complete linearly independent basis spanning the total spin 0 Hilbert space of a chain of spin-1/2 particles.\cite{rum32}  We denote the singlet projection operator acting on neighboring sites $i$ and $i+1$ by $\Pi_i^0$, which, for spin-1/2 particles, can be expressed as 
\begin{equation}
\Pi_i^0 = \frac{1}{4} - {\vec S}_i \cdot {\vec S}_{i+1},
\end{equation} 
where ${\vec S}_i$ is a spin-1/2 operator,(here and throughout $\hbar = 1$).  Figure \ref{vb}(b) shows two representative examples of $\Pi_i^0$ acting on a non-crossing valence-bond state. For spin-1/2 particles, the parameter $d$ appearing in Fig. \ref{vb}(b) is equal to 2; however, in principle, $d$ can take any value, (of course if $d\ne 2$ the Hilbert space no longer describes spin-1/2 particles). 

Of particular interest are the cases 
\begin{equation}
d = 2 \cos\frac{\pi}{k+2},
\end{equation} 
where $k$ is a positive integer.\cite{kau94}  For these values of $d$, when $k$ is finite, the non-crossing states are no longer linearly independent and the Hilbert space dimensionality of $N$ sites can be shown to grow asymptotically as $d^N$ with $d<2$. The $k \rightarrow \infty$ limit then corresponds to the case of ordinary spin-1/2 particles with $d=2$ for which the Hilbert space dimensionality grows as $2^N$.

One consequence of the reduced Hilbert space dimensionality for finite integer $k$ is that it changes the entanglement entropy associated with a valence bond.  The entanglement entropy of a subsystem $A$ of a larger system consisting of parts $A$ and $B$ is defined to be the von Neumann entropy, $S^{\rm vN}$, of the reduced density matrix $\rho_A$ obtained by tracing out the degrees of freedom in region $B$, thus
\begin{equation}
S^{\rm vN} = -{\rm Tr}[\rho_A \log_2 \rho_A].
\end{equation}
With this definition, an ordinary singlet formed by two spin-1/2 particles, with one spin in region $A$ and the other in region $B$, will have $S^{\rm vN} = 1$. However, when $d=2\cos\frac{\pi}{k+2}$,  it was shown in Ref.~\onlinecite{bon07} that if there are $M$ valence bonds connecting sites in region $A$ with sites in region $B$, then, in the $M \gg 1$ limit, because the dimensionality of the traced out Hilbert space grows as $d^M$, $S^{\rm vN} \simeq M \log_2 d$ and the entanglement per bond is $\log_2 d$.

The class of Hamiltonians studied here are all characterized by the parameter $d$ and have the form
\begin{eqnarray}
H = -\sum_i J_i \Pi_i^0, \label{model}
\end{eqnarray}
with $J_i > 0$.  For $d=2$ these models correspond to spin-1/2 AFM Heisenberg chains with $J_i$ equal to the exchange energy associated with spins $i$ and $i+1$.  For general $d$, if the $J_i$'s are uniform ($J_i = J$) the Hamiltonians (\ref{model}) can be viewed as 1+1 dimensional quantum Potts models obtained by taking the asymmetric limit of the transfer matrix of the $Q$-state Potts models\cite{martinbook} with $Q = d^2$. 

For $d \le 2$ the uniform models are all gapless, and for the special values $d= 2\cos\frac{\pi}{k+2}$ they correspond to a sequence of conformally invariant Andrews-Baxter-Forrester\cite{abf} (ABF)  models with central charges $c_k = 1-6/(k+1)(k+2)$.\cite{huse84} Physically, these ABF models can be thought of as describing chains of interacting non-Abelian particles described by $su(2)_k$ Chern-Simons-Witten theory, believed to be relevant for certain quantum Hall states.\cite{fei07}  Two special cases of these models are $k=2$ ($d=\sqrt{2}$) which corresponds to the critical 1D transverse field Ising model and $k=3$ ($d=\phi$ where $\phi = (\sqrt{5}+1)/2$ is the golden mean) which corresponds to the so-called golden chain made up of interacting Fibonacci anyons.\cite{fei07} The known universal entanglement scaling of conformally invariant 1+1 dimensional systems\cite{vid03} implies that the entanglement entropy of a block of $L$ contiguous sites, $S^{\rm vN}_L$, in the ground states of these models will scale logarithmically for $L \gg 1$ as\cite{fei07}
\begin{equation}
S^{\rm vN}_L \simeq \frac{c_k}{3} \log_2 L.
\end{equation}

When the $J_i$'s are random, the Hamiltonians (\ref{model}) can no longer be solved exactly.  However, the RSRG approach of Fisher\cite{fis94} can be straightforwardly applied for all $d \ge \sqrt{2}$ with the result that the ground states all flow to the same infinite randomness fixed point\cite{bon07,footnote1} --- one for which the bond strength distribution is the same as that of the fixed point of the random Heisenberg chain.\cite{fis94} For this fixed point, Refael and Moore\cite{ref04} have shown that if $n_L$ is the number of valence bonds leaving a given block of size $L$ (see Fig. \ref{vb}(c)), then, in the $L \gg 1$ limit, 
\begin{equation}
\overline{n_L} \simeq \frac{\ln L}{3} \simeq \frac{\ln 2}{3} \log_2 L, 
\end{equation}
where the overbar denotes a disorder average over random singlet states produced by the RSRG. This logarithmic scaling is a direct consequence of the inverse-square distribution of valence bond lengths characteristic of random singlet states.\cite{hoy07} Multiplying $\overline{n_L}$ by the entanglement per bond of $\log_2 d$ then yields the RSRG result for the asymptotic scaling of the entanglement entropy for the random ABF models, which is again logarithmic and has the form\cite{ref04,bon07}
\begin{equation}
\label{eq:SvnDIS}
S^{\rm vN}_L \simeq \overline{n_L}\ \log_2 d \simeq \frac{\ln d}{3} \log_2 L.
\end{equation}

\section{Valence-bond Monte Carlo}\label{sec:vbmc}

When applying the VBMC method\cite{san05} to Hamiltonians of the form (\ref{model}) the ground state is projected out by repeatedly applying $-H$ to a particular non-crossing valence-bond state $|S\rangle$. The result of this projection after $n$ iterations is,
\begin{eqnarray}
(-H)^n |S\rangle = \sum_{i_1,\cdots,i_n} J_{i_1} \cdots J_{i_n}
\Pi_{i_1}^0 \cdots \Pi_{i_n}^0 | S \rangle.
\end{eqnarray}
The properties of the projection operators shown in Fig. \ref{vb}(b) imply that $\Pi_{i_1}^0 \cdots \Pi_{i_n}^0 | S \rangle = \lambda_{i_1,\cdots,i_n} |\alpha\rangle$ where $|\alpha\rangle$ is a non-crossing valence-bond state with the same norm as $|S\rangle$.  The coefficient is given by $\lambda_{i_1,\cdots,i_n} = d^{-m}$ where $m$ is the number of times a projection operator acts on two sites which are not connected by a valence bond when projecting $|S\rangle$ onto $|\alpha\rangle$.\cite{san05,tra10} This projection thus leads to an expression for the ground state $|\psi\rangle$ which becomes exact in the limit of large $n$ (in our simulations we find it is sufficient to take $n = 60 N$ where $N$ is the number of sites) and has the form
\begin{eqnarray}
|\psi\rangle = \sum_\alpha w(\alpha) |\alpha\rangle,\label{wf}
\end{eqnarray}
where $w(\alpha) = J_{i_1} \cdots J_{i_n} \lambda_{i_1 \cdots i_n}$. In the simplest form of VBMC the valence-bond states $|\alpha\rangle$ contributing to $|\psi\rangle$ are sampled with probability $w(\alpha)$ by updating the sequence of projection operators $(i_1,\cdots,i_n)$ using the usual Metropolis method.\cite{san05}

To use VBMC to calculate the quantum mechanical expectation value of given operator $O$, i.e. $\langle \psi|O|\psi\rangle/\langle \psi | \psi \rangle$, it is necessary to project the ground state out of both the bra and ket states, in which case one samples from ``loop'' configurations corresponding to the valence-bond state overlaps $\langle \alpha | \beta \rangle$ with probabilities weighted by $w(\alpha)w(\beta)$.\cite{san05}  However, using the ``one-way'' VBMC described above in which one simply samples from the valence-bond basis it is possible to calculate a number of interesting quantities which can be used to characterize the intuitive valence-bond description of the ground state wavefunction.

In particular, given an observable $O$ with expectation values $O(\alpha) = \langle \alpha |O| \alpha \rangle/\langle \alpha|\alpha \rangle$ in the non-crossing valence-bond states $|\alpha\rangle$, VBMC can be used to compute the average
\begin{equation}
\langle O \rangle = \frac{\sum_\alpha w(\alpha) O(\alpha)}{\sum_\alpha w(\alpha)},
\end{equation}
for any state $|\psi\rangle$ of the form (\ref{wf}), provided $w(\alpha) \ge 0$.  In what follows, angle brackets will always denote this average, though it should be noted that $\langle O \rangle$ will in general not be equal to $\langle \psi|  O |\psi\rangle/\langle \psi| \psi \rangle$, both because the valence-bond states are nonorthogonal and because the weight factors $w(\alpha)$ are amplitudes and not probabilities. 

\section{Valence-bond entanglement}\label{sec:vbee}

One quantity that can be calculated naturally by VBMC is the valence-bond entanglement entropy, $S^{\rm VB}_L$, which, for the uniform Heisenberg chain, is defined to be equal to $\langle n_L \rangle$, the average number of valence bonds leaving a block of size $L$.\cite{alet07,chh07} To generalize $S^{\rm VB}_L$ to the ABF models with $d=2\cos \frac{\pi}{k+2}$ it is natural to multiply $\langle n_L \rangle$ by the asymptotic entanglement per bond of $\log_2 d$. For this choice, provided $n_L \gg 1$, $S^{\rm VB}_L$ will be equal to $S^{\rm vN}_L$ for any single valence-bond state.  We therefore take
\begin{equation}
S^{\rm VB}_L = \langle n_L \rangle \log_2 d.
\end{equation}

While $S^{\rm VB}_L$ is easy to compute numerically by VBMC, for a general superposition of valence-bond states it will not be equal to $S^{\rm vN}_L$.  Nonetheless, VBMC simulations\cite{alet07,chh07,kal09} of the uniform AFM Heisenberg chain with $N \simeq 100$ spins have shown numerically that $S^{\rm VB}_L$ grows logarithmically with $L$, in the same fashion as the von Neumann entanglement $S^{\rm vN}_L$. To characterize this log scaling it is convenient to introduce an effective valence-bond central charge, $c^{\rm VB}$, defined so that 
\begin{equation}
S_L^{\rm VB} \simeq \frac{c^{\rm VB}}{3} \log_2 L,
\end{equation}
in the limit $L \gg 1$. 

\begin{figure}[t]
\begin{center}
\includegraphics[width=8.5cm]{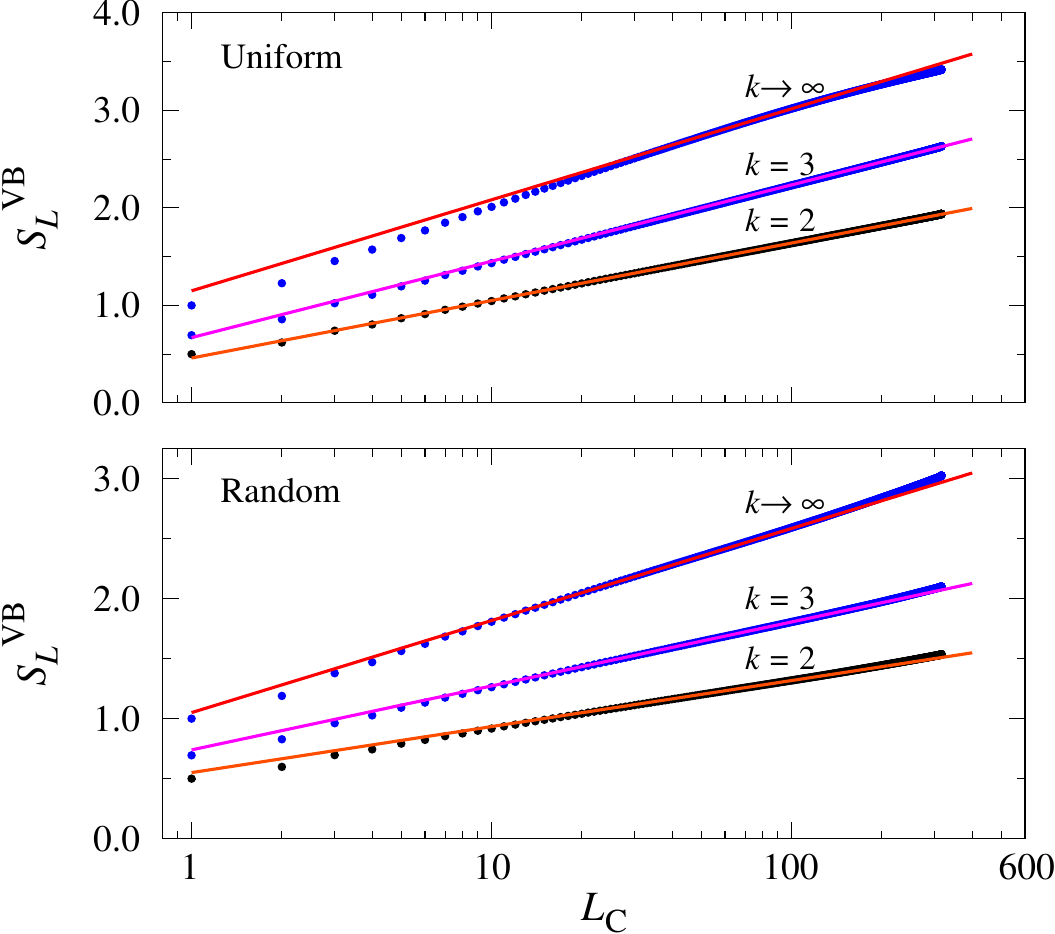}
\caption{(Color online) Log-linear plots of valence-bond entanglement $S^{\rm VB}_L$ as a function of conformal distance $L_C =(N/\pi) \sin (L \pi/N)$ for uniform (upper panel) and random (lower panel) models with $k=2,3$ and $\infty$.  For uniform models the solid lines correspond to the exact asymptotic scaling which follows from Ref.~\onlinecite{jac08}.  For random models the solid lines show the asymptotic scaling predicted by the RSRG.\cite{ref04}  Results are for periodic chains with $N = 1024$ sites.  For random chains we take the disorder strength to be $u=1$ (see Eq.~(\ref{bld})) and results are self-averaged over all blocks for 50 disorder samples.}
\label{uent}
\end{center}
\end{figure}

In addition to showing log scaling of $S^{\rm VB}_L$, previous VBMC simulations of the uniform AFM spin-1/2 Heisenberg chain have given results consistent with $c^{\rm VB}$ being close to,\cite{chh07} or even possibly equal to,\cite{alet07} $1$, the value of the true central charge for the uniform $d=2$ model.  However, Jacobsen and Saleur\cite{jac08} were able to determine the exact asymptotic scaling of $\langle n_L \rangle$ analytically for all $d \le 2$.  Their results both confirmed the log scaling of $S^{\rm VB}_L$ for $L \gg 1$ and provided an analytic result for the coefficient of the log, which yields the following expression for the valence-bond central charge,
\begin{eqnarray}
c^{\rm VB} =\frac{6\ln d}{\pi} \sqrt{\frac{2+d}{2-d}} \frac{\arccos(d/2)}{\pi-\arccos(d/2)}.
\label{cvb}
\end{eqnarray}
In the limit $d \rightarrow 2$, this expression gives $c^{\rm VB} =  12 \ln 2/\pi^2 \simeq 0.843$, which is therefore not equal to the true central charge of 1.

\begin{figure}
\begin{center}
\includegraphics[width=8cm]{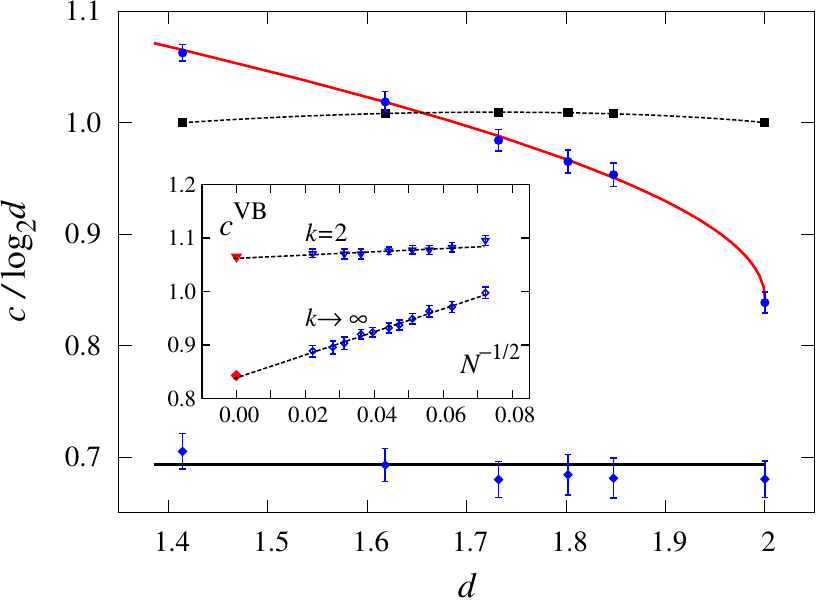}
\caption{(Color online) $d$ dependence of the valence-bond central charge, $c^{\rm VB}$, for uniform and random chains, along with the true central charges for the conformally invariant uniform models with $d = 2 \cos \frac{\pi}{k+2}$, in units of $\log_2 d$ --- the entanglement per bond.  The solid red line is the exact result of Jacobsen and Saleur \cite{jac08} for $\langle n_L \rangle$ (which equals $c^{\rm VB}/\log_2 d$ when $d=2\cos\frac{\pi}{k+2}$) vs. $d$ for uniform models, the black squares are the central charges of the ABF models for different values of $k$ (the dashed line is a guide to the eye), the black line is the RSRG result $c^{\rm VB}/\log_2 d = \ln 2$ for the random models, and the blue symbols are our VBMC results for uniform and random models. The inset shows the finite size extrapolation used to find $c^{\rm VB}$ for the uniform models with $k \rightarrow \infty$ and $k=2$.  The extrapolation clearly show the strong finite size effects for the $k\rightarrow \infty$ case with $d=2$.}
  \label{ccharge}
   \end{center}
\end{figure}

Figure \ref{uent}(a) shows our VBMC results for $S^{\rm VB}_L$ for $k=2, 3$ and $\infty$ (corresponding to $d = \sqrt{2}, \phi$ and 2, respectively) for periodic systems with $N = 1024$ sites. To minimize finite size effects when $L$ is comparable to $N/2$, $S_L^{\rm VB}$ is plotted as a function of the conformal distance $L_C =(N/\pi)\sin(\pi L/N)$. The solid lines show the exact asymptotic scaling found by Jacobsen and Saleur\cite{jac08} which clearly agree with our numerical results for $L \gg 1$.  Note that for the case $k \rightarrow \infty$ it is necessary to consider fairly large values of $L$ before entering the scaling regime, whereas for $k=2$ and 3 the scaling begins at relatively small $L$.  This fact may account for the initial numerical difficulty in determining $c^{\rm VB}$ for $d=2$ using small systems (see, however, Ref.	\onlinecite{kal09}). Presumably, the reason that the finite size effects become more pronounced as $d$ approaches 2 is because this is a critical value (for $d > 2$ the uniform models acquire a gap\cite{martinbook}).

For random Heisenberg chains $S^{\rm VB}_L$ was first computed numerically by Alet {\it et al.}\cite{alet07}.  Following the same procedure as these authors, we compute $S^{\rm VB}_L$ by determining $\langle n_L \rangle$ for particular realizations of disorder and then disorder averaging.  Throughout this paper we assume the random models are characterized by a flat bond strength distribution centered around $J=1$ of width $u$,
\begin{eqnarray}
P(J) = \frac{1}{2u}\Theta((1+u)-J) \Theta (J - (1-u)),
\label{bld}
\end{eqnarray}
where $u$ is a measure of disorder strength.  For the random ABF models we again multiply $\langle n_L \rangle$ by the entanglement per bond, and thus take
\begin{equation}
S^{\rm VB}_L =  \overline{\langle n_L \rangle}\ \log_2 d,
\end{equation}
(here again the overbar denotes disorder average). Figure \ref{uent}(b) shows log plots of our results for $S^{\rm VB}_L$ for random chains with strong disorder ($u=1$), again for $k=2,3$ and $\infty$ and $N = 1024$. The solid lines show the scaling predictions based on the RSRG\cite{ref04,bon07} for $S^{\rm vN}_L$ which clearly agree with our numerical results.  As pointed out by Alet et al.\cite{alet07}, the fact that $S^{\rm VB}_L$ and $S^{\rm vN}_L$ show the same scaling for $L \gg 1$ is to be expected if, as predicted by the RSRG, on long length scales the ground states of the random models are dominated by a single valence-bond configuration.

The log scalings of $S^{\rm VB}_L$ shown in Fig.~\ref{uent}(a) and \ref{uent}(b) are summarized in Fig.~\ref{ccharge}, which shows our VBMC results for $c^{\rm VB}$ for both uniform and random models and various values of $d$ corresponding to $k\to\infty$ and $k=2,3,4,5,6$. For the uniform models, Fig.~\ref{ccharge} also shows the exact values of $c^{\rm VB}$ (see (\ref{cvb})) which follow from the analytic results of Jacobsen and Saleur\cite{jac08} as well as the true central charges $c_k$ of the ABF models with $d=2\cos\frac{\pi}{k+2}$.\cite{footnote2} For the random models the $d$ dependence of $c^{\rm VB}$ is seen to be entirely due to the entanglement per bond, reflecting the fact that the valence bond length distribution, which determines the coefficient in front of the log scaling of $\overline{\langle n_L \rangle}$, and which depends on $d$ for the uniform case, becomes independent of $d$ when disorder is included.

\section{Valence-bond fluctuations}\label{sec:vbf}

The expectation that for random models the $L \gg 1$ scaling of $S^{\rm VB}_L$ should be the same as that of $S^{\rm vN}_L$ is based on the assumption that the valence bonds in the ground state of the model lock into a particular random singlet configuration on long length scales. This assumption is in turn based on the RSRG approach which, although it can be shown to capture the long distance properties of the fixed point exactly,\cite{fis94} is still an approximate method. Consequently, it is clearly desirable to have a direct numerical demonstration that the valence bonds are indeed locking into a particular random singlet configuration on long length scales.

To provide such a demonstration, we calculate the fluctuations in $n_L$, a quantity we refer to as the {\it valence-bond fluctuations}. To be precise, we first compute the quantity $\langle n_L^2 \rangle - \langle n_L \rangle^2$ for a particular block of size $L$ and a particular realization of disorder, and then perform a disorder average. The quantity we compute is thus
\begin{equation}
\sigma^2_L = \overline{\langle n_L^2 \rangle - \langle n_L \rangle^2}.
\end{equation}
For this choice of averaging $\sigma^2_L$ has the property that, in an idealized random singlet phase for which the ground state is precisely a single non-crossing valence-bond state, $\sigma^2_L$ would vanish, even though the number of bonds leaving a given block would be different for different realizations of disorder.

For the uniform models with $d \le 2$, Jacobsen and Saleur\cite{jac08} have also determined the asymptotic scaling of $\sigma^2_L$ (in this case there is, of course, no disorder average). Like $\langle n_L \rangle$, $\sigma_L^2$ scales logarithmically with $L$ for $L \gg 1$, with
\begin{eqnarray}
\sigma_L^2 \simeq b\ln L,
\end{eqnarray}  
and the analytic results of Ref.~\onlinecite{jac08} can again be used to obtain an exact analytic results for the coefficient, $b$, as a function of $d$,
\begin{eqnarray}
b =\frac{4}{\pi}\sqrt{\frac{2+d}{(2-d)^3}}\frac{2\arccos(d/2)-\sqrt{4-d^2}}
{\pi-\arccos(d/2)}.
\end{eqnarray}

\begin{figure}[t]
 \begin{center}
    \includegraphics[width=8.5cm]{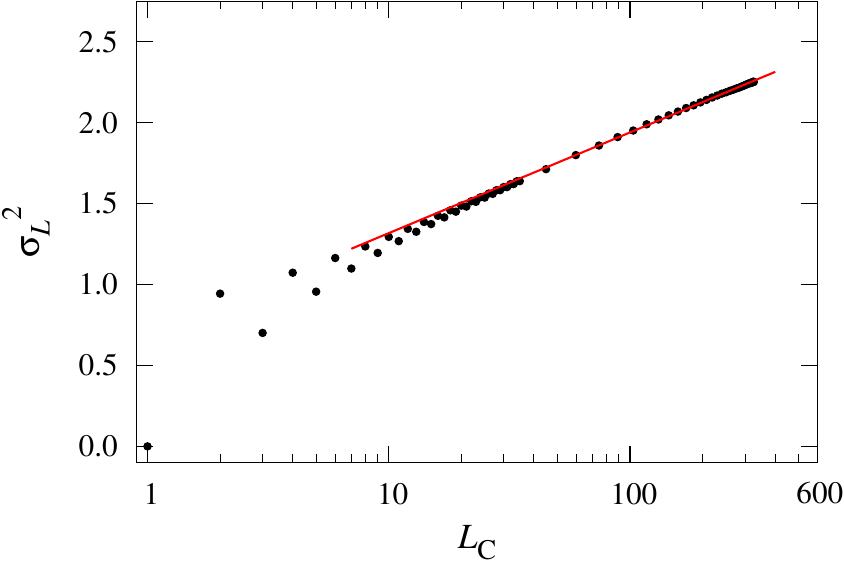}
  \caption{(Color online) Log-linear plot of the valence-bond fluctuations $\sigma_L^2$ as functions of the conformal distance $L_C =(N/\pi) \sin (L \pi/N)$ for the case $k\to\infty$. The exact log scaling obtained by Jacobsen and Saleur is shown by the solid line. For small $L$ a strong even/odd effect --- explained in the text --- can be seen.  Results are for periodic chains with $N = 1024$ sites.}
  \label{fluct-unf}
  \end{center}
\end{figure}

Figure~\ref{fluct-unf} shows a log-linear plot of our VBMC results for $\sigma_L^2$ calculated for the uniform model with $k\to\infty$. A line corresponding to the exact asymptotic log scaling found by Jacobsen and Saleur\cite{jac08} is also shown. It is readily seen that our numerical results agree well with the predicted asymptotic scaling, and can be regarded as further numerical confirmation of the field-theoretic analysis of Ref.~\onlinecite{jac08}. We note that the log scaling of $\sigma_L^2$ directly demonstrates the existence of bond fluctuations on all length scales in the ground state of the uniform model.

\begin{figure}[t]
 \begin{center}
    \includegraphics[width=8.5cm]{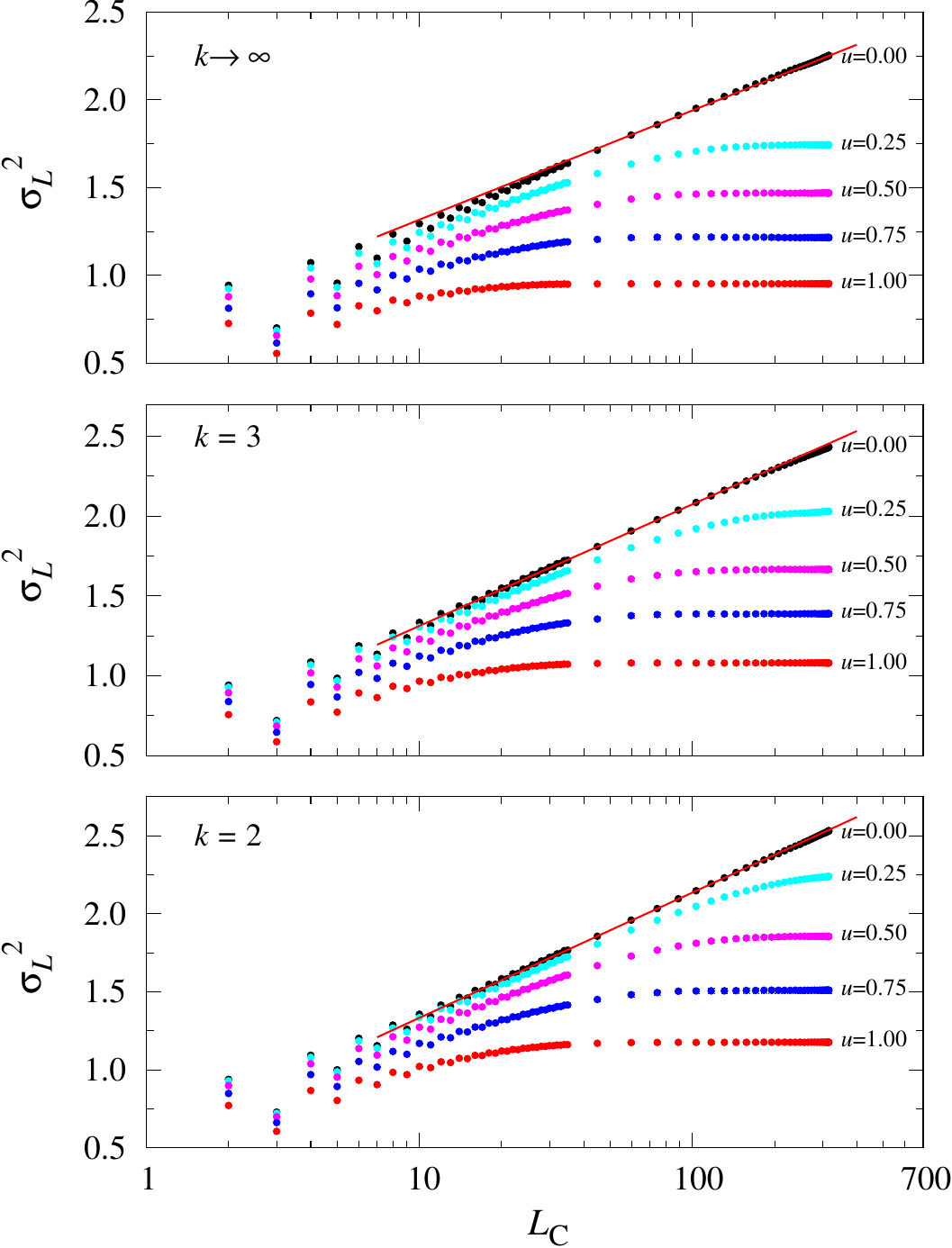}
  \caption{(Color online) Log-linear plots of valence-bond fluctuations as a function of conformal distance $L_C=(N/\pi) \sin (L \pi/N)$, for both uniform and random chains for $k\rightarrow \infty$ (upper panel), $k=3$ (middle panel) and $k=2$ (lower panel).  For uniform chains these fluctuations grow logarithmically with $L_C$ in agreement with the analytic results of Ref.~\onlinecite{jac08} (solid red lines), indicating that bonds are strongly fluctuating on all length scales. For random chains the fluctuations saturate beyond a given finite length scale $\xi$, signaling the formation of a random singlet phase in which the bonds have locked into a particular random singlet configuration on long length scales. The parameter $u$ (defined in the text) is a measure of the disorder strength and the results clearly show that the saturation length scale grows with decreasing $u$. Results are for periodic chains with $N = 1024$ sites and for random models are self-averaged over all blocks for 100 disorder samples.}
  \label{fluc}
  \end{center}
\end{figure}

Figure \ref{fluct-unf} also shows that for small $L$ the valence-bond fluctuations $\sigma_L^2$ oscillate strongly as the block length $L$ changes from even to odd. The origin of this even/odd effect can be understood by first considering a state in which the bonds are all of length $1$ (i.e. a dimerized state).  In this case there would be two ground states corresponding to the two distinct dimerizations and the translationally invariant ground state would be an equal superposition of these two dimerized states.  One can readily check that in such a state $\sigma^2_L = 1$ when $L$ is even, and $\sigma^2_L = 0$ when $L$ is odd.  We believe that the even/odd oscillations apparent in Fig.~\ref{fluct-unf} for small $L$ are due to the significant contribution of such dimerized regions (at least on small length scales) to the ground state wavefunction.

For random models, the RSRG approach\cite{fis94} shows that on long length scales the bonds lock into a random singlet state.  At the same time, on short length scales (if disorder is weak) it is natural to expect that the bonds will fluctuate strongly, as they do in the uniform models. This implies the existence of crossover length scale $\xi$ which characterizes the transition from the uniform regime to the random-singlet regime of these models with increasing $L$. One can then expect the valence-bond fluctuations $\sigma^2_L$ to not differ much from their value for the uniform models when $L \ll \xi$, but for $L \gg \xi$ the fluctuations $\sigma^2_L$ should {\it saturate}. This saturation is due to the fact that, once the block size $L$ becomes much larger than the crossover length $\xi$, the bond fluctuations occurring outside of a distance $\xi$ from the two boundaries of the block will not change the number of bonds leaving the block, and hence will not contribute to the valence-bond fluctuations.

\begin{figure}[t]
 \begin{center}
    \includegraphics[width=8.5cm]{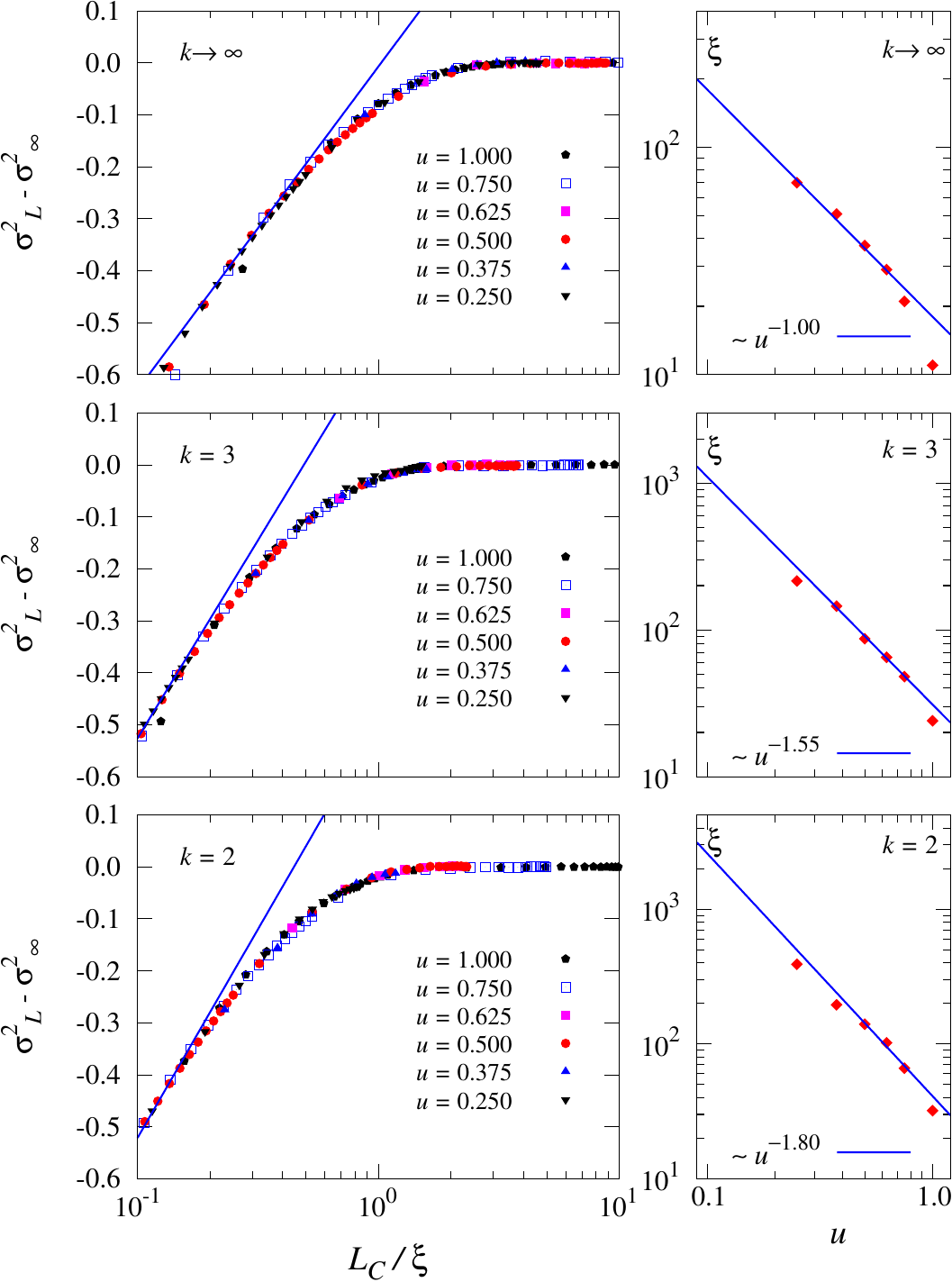}
  \caption{(Color online) (Left) Scaling plots of the valence-bond fluctuations (with the saturation value $\sigma^2_\infty$ subtracted out),  $\sigma_{L}^2-\sigma^2_\infty$, for $k\to\infty$, $k=3$, and $k=2$ and various disorder strengths $u$.  In these plots the $x$-axes are rescaled by choosing the disorder-dependent crossover length $\xi$ so that the data for different $u$ collapse onto a single curve when plotted vs. $L_C/\xi$ where $L_C=(N/\pi) \sin (L \pi/N)$. The solid lines show the predicted log scaling for the uniform models.\cite{jac08} The values of $\xi$ determined by this scaling procedure are summarized on Table \ref{table:xi}.    (Right) Log-log plots of the crossover length scale $\xi$  as a function of $u$ for $k\to\infty$, $k=3$, and $k=2$. Solid lines represent the power-law scaling determined from the data: $\xi \sim u^{-1}$ for $k\to\infty$, $\xi \sim u^{-1.55}$ for $k=3$, and $\xi \sim u^{-1.8}$ for $k=2$.}
  \label{scaling}
  \end{center}
\end{figure}

Figure \ref{fluc} shows log-linear plots of our results for $\sigma^2_L$ for the case $k\rightarrow \infty$ (corresponding to the Heisenberg chain, with $d=2$), $k=3$ (corresponding to the golden chain, with $d=\phi$) and $k=2$ (corresponding to the critical transverse field Ising model, with $d = \sqrt{2}$) for both uniform and random models. For the random models the $J_i$'s are taken to be distributed according to (\ref{bld}), where $u$ is a measure of disorder strength. It can be observed from Fig.~\ref{fluc} that the valence-bond fluctuations $\sigma^2_L$ for the random models saturates, regardless of how weak the disorder is, on a finite length scale $\xi$ which grows as $u$ decreases.

The observation of this saturation, which indicates a finite crossover length scale $\xi$ beyond which the valence bonds effectively lock into a random singlet configuration, together with the log scaling of $\overline{\langle n_L \rangle}$, which indicates a power-law distribution of valence bond lengths, provides a direct numerical proof of random singlet phase formation in these models. 

\section{Crossover length scale}\label{sec:clength}

As described in the previous section, the saturation of $\sigma_L^2$ with increasing $L$ for disorder of any strength $u$ implies the existence of a finite fluctuation length scale $\xi$ which characterizes the transition from the resonating regime with $L \ll \xi$ to the saturation regime with $L \gg \xi$. This length scale $\xi$ is essentially the crossover length scale from the uniform regime to the disordered regime, which has been studied in the literature both analytically and numerically for a number of models.\cite{dot92,gia88,laf04a,laf04} 

For $k\to\infty$ and $k=2$ analytic results for the dependence of $\xi$ on $u$ can be obtained for the case of weak disorder\cite{dot92} by mapping the models (\ref{model}) onto disordered Luttinger liquids.\cite{gia88}   For $k\to\infty$ the model (\ref{model}) corresponds to an isotropic spin-1/2 Heisenberg chain which, via a Jordan-Wigner transformation, can be mapped onto a 1D interacting spinless Fermi gas with a particular interaction strength.  Similarly, for the case $k=2$ the model (\ref{model}) corresponds to the 1D transverse field Ising model, and a pair of independent but identical 1D transverse field Ising models can be mapped onto a spin-1/2 XX model which can in turn be mapped onto a (in this case free) 1D spinless Fermi gas.  The resulting predictions\cite{dot92,gia88} for the scaling of the crossover length scale $\xi$ with disorder strength $u$ for these two cases are that $\xi \sim u^{-1}$ for $k\to\infty$ and $\xi \sim u^{-2}$ for $k=2$.  Numerical results for $\xi$, based on scaling analyses of the spin-spin correlation function\cite{laf04a,laf04} and the spin stiffness\cite{laf04} of the Heisenberg chain (using quantum Monte Carlo) and the XX chain (by exact diagonalization) have shown results consistent with these weak disorder renormalization group predictions.  

\begin{table}[t]
\begin{center}
\caption{Crossover length scale $\xi$ extracted from the scaling analysis of the valence-bond fluctuations $\sigma_{L}^2$ for different disorder strengths $u$ for $k\to\infty$, $k=3$, $k=2$.} \label{table:xi}
\begin{tabular}{|c|c|c|c|}
	\hline
$u$ &  $k\to\infty$ & $k=3$ & $k=2$ \\
	\hline
0.250   & 70  & 215 & 390\\
0.375  & 51  & 145 & 195\\
0.500  & 37  & 87 &140\\
0.625  & 29  & 65 &102\\
0.750  & 21  & 48 &66\\
1.000  & 11 & 24 & 32\\
\hline
\end{tabular}
\end{center}
\end{table}

It is possible to determine the dependence of $\xi$ on $u$ by performing a scaling analysis of the valence-bond fluctuations $\sigma_L^2$.  To do this, we first subtract the large $L$ saturated value of the fluctuations ($\lim_{L\rightarrow\infty} \sigma^2_{L} = \sigma^2_\infty$) obtained by extrapolating the data shown in Fig.~\ref{fluc} and attempt to collapse the data by assuming a scaling function $f$ and a $u$ dependent $\xi$ for which 
\begin{equation}
\label{scaling_funct}
\sigma_L^2(u)-\sigma_\infty^2(u)=f\left(\frac{L_C}{\xi(u)}\right).
\end{equation}

For each value of $u$, $\xi(u)$ is chosen so that data for $\sigma_L^2-\sigma_\infty^2$ collapse onto a single curve, with the center of the crossover regime being $L_C/\xi \simeq 1$.   Note that to avoid the even-odd effect pointed out earlier for small $L$ we only use odd values of $L$, and to minimize finite size effects for large $L$ we use the conformal distance $L_C$ in the scaling analysis.

Results of carrying out this analysis for the cases $k\to\infty$, $k=3$, and $k=2$ are shown on the left side of Fig.~\ref{scaling}. The VBMC results for $\sigma_L^2-\sigma_\infty^2$ can be seen to be well collapsed onto a particular scaling function $f(L_C/\xi)$, according to the definition (\ref{scaling_funct}). The values we obtain for $\xi$ for these models corresponding to various disorder strength $u$ are given in Table \ref{table:xi}. 

On the right side of Fig.~\ref{scaling}, log-log plots of the crossover length scales $\xi$ as a function of the disorder strength $u$ are shown. The exponents characterizing the divergence of $\xi$ are determined by fitting the data to the power law $u^{-\eta}$ (solid lines). For $k\to\infty$, we find $\eta \simeq 1$, consistent with the weak disorder renormalization group prediction\cite{dot92,gia88} of $\eta = 1$.  For $k=2$, we find $\eta \simeq 1.8$, which is somewhat less than the predicted value of $\eta = 2$.  One possible reason for the poorer agreement in this case is that for $k=2$ the length scale $\xi$ is significantly larger than for $k \rightarrow \infty$ for a given disorder strength, and it may therefore be necessary to study larger system sizes in order to enter the scaling regime for the valence-bond fluctuations.  

For the case $k=3$, for which the model (\ref{model}) corresponds to a disordered golden chain, we find the exponent $\eta \simeq 1.55$. Note that for this system (with $d = \phi$), and, in fact, for all cases for which $d \ne 2,\sqrt{2}$, there is no simple mapping of (\ref{model}) to a disordered 1D Luttinger liquid.  It is therefore not possible to apply  the same weak disorder renormalization group analysis to these models that can be used to obtain the exponent $\eta$ for $d=2$ and $d=\sqrt{2}$.  To the best of our knowledge there are no known analytic results for $\eta$ or for these more general models and we believe our result for $k=3$ represents the first numerical calculation of such an exponent.

\section{Conclusions}\label{sec:conclusion}

In this paper we have presented the results of a VBMC study of both uniform and random Hamiltonians of the form (\ref{model}). Both the valence-bond entanglement entropy $S_L^{\rm VB}$ and the valence-bond fluctuations $\sigma^2_L$ were calculated for these models.  For uniform models both these quantities were found to scale logarithmically with $L$ and our results agreed well with analytic results obtained through a field-theoretic analysis by Jacobsen and Saleur.\cite{jac08} For random models $S_L^{\rm VB}$ was also found to scale logarithmically with $L$, consistent with predictions based on the RSRG,\cite{ref04,bon07} while $\sigma^2_L$ was found to saturate once $L$ exceeded a disorder dependent crossover length scale $\xi$, signaling the expected locking of the valence bonds into a particular random singlet configuration on long length scales.

By performing a scaling analysis of the valence-bond fluctuations we were able to determine the dependence of $\xi$ on disorder strength. For the cases $k\to\infty$ (spin-1/2 Heisenberg model) and $k=2$ (transverse field Ising model) our results were consistent with those based on a weak disorder renormalization group approach\cite{dot92,gia88} as well as previous numerical work,\cite{laf04,laf04a} although for the case $k=2$ we may not have fully entered the scaling regime.  An appealing feature of our bond fluctuation based approach is that it can be used to determine $\xi$ for any value of $k$, not just $k\rightarrow \infty$ and $k=2$ for which the models (\ref{model}) can be mapped onto 1D Luttinger liquids (the starting point for the weak disorder renormalization group approach). For example, we have determined for the first time the crossover length scale $\xi$ and the corresponding exponent $\eta \simeq 1.55$ for the model (\ref{model}) with $k=3$, which corresponds to the disordered golden chain.

\acknowledgments The authors acknowledge US DOE Grant No. DE-FG02-97ER45639 for support. Useful discussions with K. Yang, A. Sandvik, P. Henelius, J. A. Hoyos, and F. Alet are sincerely acknowledged. Computational work was performed at the Florida State University High Performance Computing Center.


\begin{thebibliography}{}
\bibitem{footnote2} It is interesting to note that while $c^{\rm VB}$ and $c_k$ are not equal for the uniform ABF models, they are close in magnitude.  This is primarily due to the fact that the entanglement per bond, $\log_2 d$, is equal to $c_k$ for $k=2$ and $k \rightarrow \infty$, and very close to $c_k$ for $k > 2$.
\end{thebibliography}
\end{document}